# Spatiotemporal Cluster Analysis of Gridded Temperature Data - A Comparison Between K-means and MiSTIC

E Ankitha Reddy*, KS Rajan**
*(LSI, International Institute of Information Technology (IIIT), Hyderabad
Email: ankitha.eravelli@research.iiit.ac.in)
** (LSI, International Institute of Information Technology (IIIT), Hyderabad
Email: rajan@iiit.ac.in)

----------------------------------------\*\*\*\*\*\*\*\*\*\*\*\*\*\*\*\*\*\*\*\*\*\*\*\*----------------------------------

## Abstract:

The Earth is a system of numerous interconnected spheres, such as the climate. Climate's global and regional influence requires understanding its evolution in space and time to improve knowledge and forecasts. Analyzing and studying decades of climate data is a data mining challenge. Cluster analysis minimizes data volumes and analyzes behavior by cluster. Understanding invariant behavior is as crucial as understanding variable behavior. Gridded data from two sources: Grided IMD data and CMIP5 HadCM3 decadal experiments, are studied using K-Means and MiSTIC clustering techniques to explore spatiotemporal clustering of maximum and minimum temperatures. The boundaries of k-means clustering correspond with topography. The Indian subcontinent's physiographic, climatic, and topographical characteristics affect MiSTIC's core areas. Both techniques yield overlapping clusters. The datasets' MiSTIC cluster counts varied significantly. The impact of data on this technique is shown in how the datasets group Himalayas.

----------------------------------------\*\*\*\*\*\*\*\*\*\*\*\*\*\*\*\*\*\*\*\*\*\*\*\*----------------------------------

## I.   INTRODUCTION

The Earth is a highly complex and dynamic system formed by synergizing phenomena in its varied spheres. One such significant phenomenon is the Earth's climate. The considerable impact of the Earth's climate on a global and regional scale raises a need to study its evolution over space and time, thus enhancing our knowledge and predictability. Analyzing the enormous amounts of observed and modeled climate data over many decades poses a data mining challenge.

Cluster analysis techniques in data mining reduce the massive amounts of data and allow cluster-wise behavior analysis of the phenomenon. The cluster analysis algorithms have been implemented in various research domains and can be classified using different criteria [1]. The existing clustering techniques are primarily spatial clustering methods that cluster the data in the spatial domain. It is essential to include the temporal dimension through spatiotemporal clustering for studying climate data. Understanding the invariant behavior of a phenomenon is as vital as variant behavior [2].

This study explores spatiotemporal clustering of maximum and minimum temperature data and requires spatially continuous and temporally consistent gridded datasets. The data comes from two data sources: 1) Gridded IMD data and 2) CMIP5 HadCM3 decadal experiments, which are studied using two methods: 1) K-Means clustering and 2) MiSTIC.

## II.    MATERIALS AND METHODS

### A.  Data

#### 1)  *Gridded IMD:*

The IMD high-resolution gridded daily temperature dataset for the Indian subcontinent was constructed using data from 395 quality-controlled stations from 1969 and 2005. The station temperature data were interpolated into $1^{o}x1^{o}$ grids and organized into 31x31 grid points using a modified version of Shepard's angular distance weighting. The error in cross-validation is less than 0.5 $^{o}$C [3]. Leap years are represented with data for 366 days. Temperature unit is $^{o}$Celsius. The gridded data for 2008 and beyond is based on real-time operational data from about 180 stations [4].

#### 2)  *IPCC CMIP5 HadCM3:*

The Hadley Centre in the United Kingdom produced HadCM3, a coupled atmosphere-ocean general circulation model (AOGCM). It was one of the primary models used in the IPCC Assessment Reports from 2001 [5]. Unlike previous AOGCMs at the Hadley Centre and elsewhere, HadCM3 does not need flux adjustment. This is mainly attributed to HadCM3's superior ocean mixing scheme, improved ocean resolution, and a strong match between the atmospheric and oceanic components. The two components of HadCM3 are the HadAM3 atmospheric model and the HadOM3 ocean model. Simulations employ a 360-day calendar with 30 days every month [5].

This research utilized temperature data from the HadAM3 portion of the HadCM3 model. HadAM3 is a grid point model with a horizontal resolution of $3.75^{o}$ x $2.5^{o}$ in longitude and latitude and 96x73 grid points on the scalar grid [5]. The minimum and maximum temperatures from 1989 to 2019 have been used [6]. In order to offer a uniform resolution for both source datasets, the gridded temperature data from HadCM3 from 1989 to 2019 has been resampled to $1^{o}x1^{o}$ spatial resolution. The temperature is expressed in $^{o}$Kelvin.

### B.  Concept of Methods

#### 1)  *K Means Clustering: A Centroid-Based Technique:*

The most straightforward and fundamental approach to clustering is partitioning, which splits a collection of items into many clusters. The number of output clusters can be provided as background information to keep the issue statement brief. This parameter defines the initial and final result of partitioning methods—the clustering iterates to maximize an objective partitioning criterion based on distance, such as a dissimilarity function. Regarding the data set properties, the function strives for high intra-cluster similarity and low inter-cluster similarity. The iterations continue until the resultant clusters are stable. In other words, the clusters generated in this iteration are identical to those formed in the preceding iteration [1].

#### 2)  *MiSTIC: Mining Spatio-Temporally Invariant Cores:*

MiSTIC is a data format generalized approach based on watershed delineation, neighborhood analysis, and frequent item mining to discover the collection of geographically dispersed areas heavily impacted by a phenomenon over time. The approach includes a spatial analysis phase to identify focal points and a spatio-temporal analysis to determine core areas or cores throughout the entire data period. Cores are categorized as Cores with Contiguous Points (CC) and Cores with a Defined Radius (CR) to comprehend their behavior, as the nature of the cores is influenced by their immediate neighborhood. The cores were further classified to assess the spatiotemporal invariance by considering the frequency of recurrence of the focus points that compose them. These include cores with highly dominating points (CHD), cores with less dominating points (CLD), and cores with no dominating points (CND). As a result, the frequently/predominantly occurring focal points encapsulate the confined invariant nature of a dynamic phenomenon. In contrast, the neighborhood restrictions capture the extent of the dynamic nature and direction of the effect of the phenomena, if any [2]

  

## III.    RESULTS AND DISCUSSION

All The methods in the study have been implemented on four datasets: Gridded IMD Tmax (tmax), Gridded IMD Tmin (tmin), HADCM3 Tmax (tasmax), and HADCM3 Tmin (tasmin).

The number of clusters for k-means clustering was set to 8, 10, and 12 clusters. The cluster map of k-means clustering can be seen in Fig.1 and Fig.2 for IMD and HadCM3 datasets.

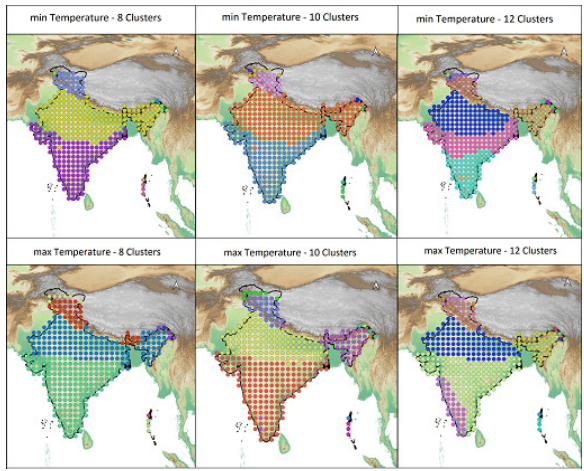

Fig. 1  K-means cluster map for Gridded IMD  minimum and maximum Temperatures

The Himalayas, Central and North-Eastern India, the Deccan plateau, and the Andaman and Nicobar Islands are the four primary clusters identified by the 8 and 10 clusters for gridded IMD minimum temperature. In contrast, the 12 clusters split the Indian regions into the trans-Himalayas, the Himalayas, the tropical plains and highlands, the northeastern region, the peninsular region, and the islands. For this dataset, the chosen cluster output is 12 clusters.

Regarding gridded IMD maximum temperature, the Himalayas, Central India, North-Eastern India, the Deccan plateau, and the Andaman and Nicobar Islands are the 4 or 5 prominent clusters by the 8 and 10 clusters, respectively. The twelve clusters divide India into the trans-Himalayas, Himalayas, Indo-Gangetic plains and desert, northeastern region, peninsular region, western coast, and islands. This dataset's chosen cluster output is 12 clusters.

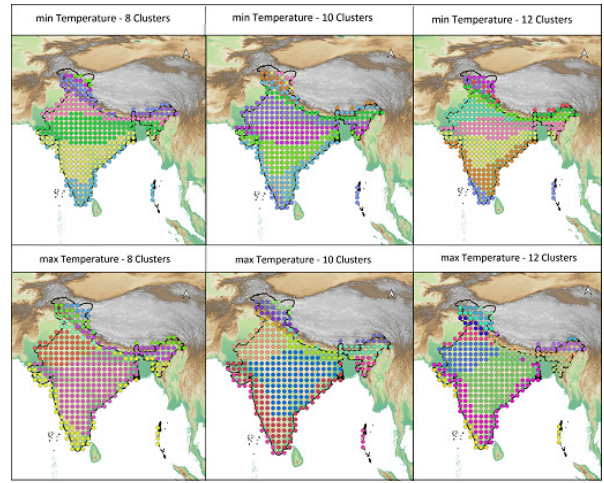

Fig. 1  K-means cluster map for HadCM3 minimum and maximum temperatures

HadCM3 min. temperature: The trans-Himalayas, the northern and eastern Himalayas, the Indo-Gangetic and desert regions, the central highlands, the Deccan plateau, the coastal region, and the islands are the eight clusters that divide the Indian region according to HadCM3's minimum temperature. It has been shown that by dividing the western and eastern ghats, and the coastal plains, ten clusters offer improved clustering. However, compared to the 10 clusters, the 12 clusters offer no more advancement. This dataset's chosen cluster output consists of 10 clusters.

HadCM3 max. Temperature: The trans-Himalayas, the Himalayas, the Indo-Gangetic plains, the desert, the central highlands, the Deccan plateau, the western coast, and the islands are the eight clusters that divide the Indian region for HadCM3 maximum temperature. It is shown that dividing the western coast, western ghats, and eastern ghats and coasts results in 10 groups that offer improved grouping. The Indo-Gangetic plains are divided into two independent clusters by the 12 clusters, while the dry and semi-arid regions of Rajasthan and Gujarat are combined into a single cluster. This dataset's chosen cluster output consists of 10 clusters.

MiSTIC needs input data dispersed spatially and organized by year and information about the 8-connected neighborhood. Each dataset's data has been averaged over each spatial location for each year to prepare the input datasets for MiSTIC. Using MiSTIC, the data was processed every year to create year-wise zones. The year-by-year zones have been processed to generate a single zone for each spatial location by identifying the zone of maximum occurrence. The clustering generated by MiSTIC for the four datasets is shown in Fig.3.

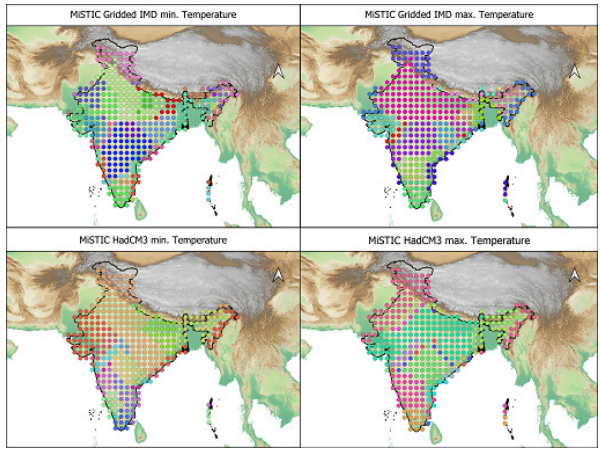

Fig. 3 MiSTIC cluster map for gridded IMD and HadCM3 minimum and maximum temperatures

A focus point is deemed frequent for this study if it has happened at the exact location for 12 years or more (i.e., 38% out of 31 years).

For gridded IMD min. temperature, The Indian area is separated into the northern and eastern Himalayas, Indo-Gangetic plains, semi-arid regions, the Indian desert, the Bengal basin, northeastern hill ranges, the central highlands, the eastern plateau, the Deccan plateau, the coastal plains, and river basins, and the islands.

The northern and eastern Himalayas, the Indo-Gangetic plains and desert regions, the Bengal basin, the northeastern hill ranges, the central highlands, the eastern plateau, the Deccan plateau, the coastal plains, and the islands are the regions of India that are divided by the MiSTIC for gridded IMD maximum temperature.

According to HadCM3's minimum temperature, the Indian region is split into the Himalayan and semi-arid regions, the Indo-Gangetic plains, the arid desert region, Bengal basin, northeastern region, Deccan plateau, coastal plains and river basins, and islands.

For HadCM3 maximum temperature, the Indian region is divided into the northern and eastern Himalayas and the Indian desert, the semi-arid regions and Indo-Gangetic plains, the Bengal basin, the Deccan plateau, the western ghats, and the western coast, the eastern coast, and the islands.

The cluster-wise elevation-slope maps plotted to visualize the trends of each cluster's mean elevation and mean slope in k-means clustering are available in Fig.4 and Fig.5.

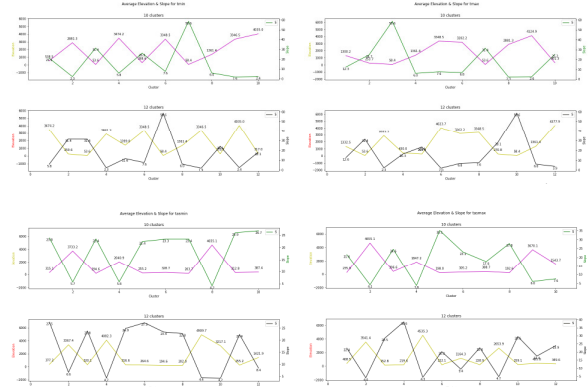

Fig. 4 Average Elevation vs Slope for K-means on four datasets

Clustering using K-means for gridded IMD data The mean elevation and mean slope trends in k-means clustering can be used to observe the inter-cluster variability of each cluster's mean elevation and mean slope. It can be observed that almost half of the clusters for k-means clustering on gridded IMD min. and max. temperatures fall within the mean elevation range of below 1500 m.

Clustering using K-means for HadCM3 data The cluster-wise mean slope and elevation plots for HadCM3 min. temperature show how closely the trends of 10 and 12 clusters resemble each other. Less than 30\% of the clusters have a mean elevation of more than 2000 m, compared to 50\% in k-means clustering for gridded IMD data.

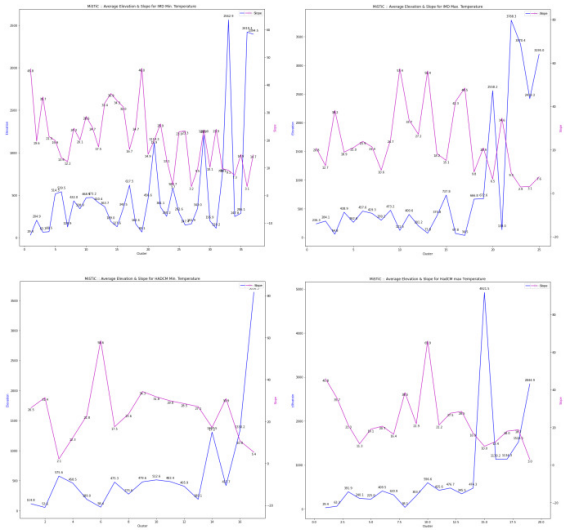

Fig. 5  Average Elevation vs Slope for MiSTIC on four datasets

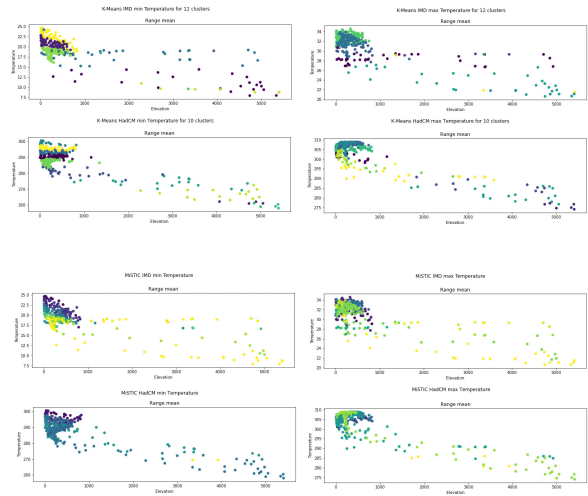

Fig. 6  Cluster-wise data distribution for four data sets using two methods

The number of clusters in MiSTIC is a process-dependent parameter rather than an input parameter, in contrast to k-means clustering. Compared to the number of clusters formed for HadCM3 datasets, it is evident that the number of clusters generated for gridded IMD datasets is substantially higher. This shows the potential of input data quality to impact the analysis. Fig.6 represents the chosen / resultant clusters' temperature range vs. elevation charts.

The k-means algorithm's clustering yielded the following classes:

For gridded IMD data: The Trans Himalayas - The Himalayas  - The Indo-Gangetic plains and desert - Northeast India - The peninsular region  - The West Coast - Andaman and Nicobar islands

For HadCM3 data: The Trans Himalayas - The northern and eastern Himalayas - Indo-Gangetic plains - The Indian desert - The central highlands - The Deccan plateau - The western ghats and the eastern ghats - The western coastal plains - Andaman and Nicobar islands

The following classes are obtained by clustering using the MiSTIC algorithm:

For gridded IMD data: The northern and eastern Himalayas - The Indian desert - Semi-arid regions - The Indo-Gangetic plains - The northeastern mountain ranges - The Bengal basin - The central highlands - The eastern plateau - The Deccan plateau - The coastal plains and river basins - Andaman and Nicobar islands

For HadCM3 data: The Himalayas and the desert region - The Indo-Gangetic plains and semi-arid regions - Northeastern India - The Bengal basin - The Deccan plateau - The western ghats and the west coast - The East Coast - Andaman and Nicobar islands

The borders of the k-means clustering observably correlate with changes in the topography. In comparison, the physiographic divisions, climatic zones, and terrain variations over the Indian subcontinent all contribute to the core regions produced by MiSTIC. Considerable overlap exists between the clusters produced by the two approaches.

The climatic, geographic, and physiographic zones of India [7] [8] [9] are used as the basis for the investigation of the clustered regions or zones in this study.

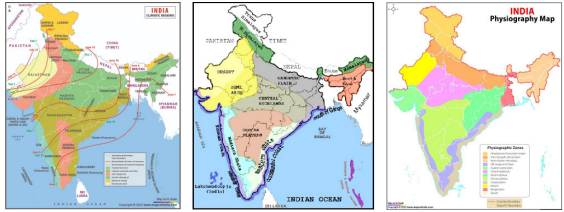

Fig. 7 Climatic, geographic, and physiographic zones of India

## IV. CONCLUSION

Even though k-means obtained a fundamental clustering of Indian areas, it ignores the spatial correlation in the data. The number of input clusters determines the number of resulting clusters. In contrast, MiSTIC uses a data-driven approach where data is the primary input. In order to determine the zones in the area, this method analyzes the data utilizing watershed delineation, neighborhood analysis, and frequent item mining techniques. The algorithm, however, analyzes data one snapshot at a time for every year throughout the relevant timeframe.

In MiSTIC, the impact of the input data on the resulting clusters is notable. The number of clusters produced for the datasets from the two sources differed significantly. Other variables, such as data continuity for better neighborhood analysis, may influence the outcome. Potential clusters could break due to the absence of data or data gaps in areas outside India but between parts of India – for example, Nepal. This will affect the outcome. In the case of IMD data, only the data for the Indian region was obtainable, whereas HadCM has global data availability. As a result, the Himalayan ranges' clustering changes between the two datasets.

## ACKNOWLEDGMENT

The authors wish to thank the Lab for Spatial Informatics, International Institute of Information Technology, Hyderabad for their prior work.